\documentclass[aps,prc,reprint,superscriptaddress,nofootinbib]{revtex4-1}
\usepackage{graphicx}
\usepackage{multirow}
\usepackage{comment}
\usepackage{color}

\usepackage{amssymb}
\usepackage{amsmath}
\usepackage{color}
\usepackage{lineno}

\def\nuc#1#2{\relax\ifmmode{}^{#1}{\protect\text{#2}}\else${}^{#1}$#2\fi}

\newcommand{\vecr}{{\vec r}}

\newcommand{\be}{\begin{eqnarray}}
\newcommand{\ee}{\end{eqnarray}}

\newcommand{\bwt}{\begin{widetext}}
\newcommand{\ewt}{\end{widetext}}

\begin{document}

\title{Post-prior equivalence for transfer reactions with complex potentials}


\author{Jin Lei}
\email[]{jinl@ohio.edu}

\affiliation{Institute of Nuclear and Particle Physics, and Department of Physics and Astronomy,
Ohio University, Athens, Ohio 45701, USA}

\author{A. M. Moro}
\email[]{moro@us.es}

\affiliation{Departamento de FAMN, Universidad de Sevilla,
Apartado 1065, 41080 Sevilla, Spain.}


\begin{abstract}
In this paper, we  address the problem of the post-prior equivalence in 
the calculation of inclusive breakup and transfer cross sections. 
For that, we employ the model proposed by Ichimura, Austern, and Vincent 
[Phys. Rev. C 32, 431 (1985)], conveniently generalized to include the 
part of the cross section corresponding the transfer to bound states. 
We pay particular attention to the case in which the unobserved particle 
is left in a bound state of the residual nucleus, in which case the 
theory prescribes the use of a complex potential, responsible for the 
spreading width of the populated single-particle states. We see that 
the introduction of this complex potential gives rise to an additional 
term in the prior cross section formula, not present in the usual case of 
real binding potentials. The equivalence is numerically tested for 
reaction induced by deuterons.
\end{abstract}


\date{\today}%
\maketitle

\section{Introduction \label{sec:intro}}
The post-prior equivalence of the transition amplitude for direct nuclear 
reactions involving different rearrangement channels is a key
result of nuclear reaction theory. This result provides two formally
equivalent ways of expressing the transition amplitude, depending on whether
the main interaction appearing in the transition operator is that based
on the initial or final internal Hamiltonians. The result holds for the
exact transition amplitude and, in the case of transfer reactions, also
for the DWBA limit. In this latter case, the post and prior expressions
are formally identical, differing only in the transition operator. This result has indeed
been confirmed in practical cases.

In the case of inclusive breakup reactions of the form $a+A \to B^*+b$,
where $a=b+x$ and $B^*$ is any $A+x$ state, the problem has deserved
attention in the past. Several groups proposed formulae for the
calculation of inclusive cross sections using either the post or prior
DWBA representations \cite{Aus81,Kas82,Ich86,Uda81}. Ichimura, Austern and
Vincent(IAV) \cite{Ich86} showed that the post and prior equivalence
holds also for these inclusive processes but, in this case, it involves
an additional term, not present in the usual transfer process between
bound states. This terms arises from the non-orthogonality between the
initial ($a+A$) and final ($b+B$) partitions. Although some authors
(see e.g. Li, Udagawa and Tamura \cite{Li84}) regarded this term as nonphysical, in
our recent work \cite{Jin15b} we showed in practical cases that the inclusion of
this term is essential to preserve the post-prior equivalence and to
reproduce correctly the experimental data.

The calculations of Ref.~\cite{Jin15b} were restricted to
unbound $x+A$ states (i.e.\ $E_x>0$, where $E_x$ is the final relative
energy between $x$ and $A$). However, the  $E_x<0$ case was not considered.
 This region would correspond to bound states of the residual $B$ system and, hence,  the process $a+A \to B+b$ becomes a transfer reaction in the usual sense. In some models, such as the DWBA, the scattering amplitude involves the overlap function between the $A$ and $B$ systems, i.e.\ $\langle A | B \rangle$. Although these overlaps should be in principle obtained from many-body wave functions of $A$ and $B$, they are most commonly approximated the single-particle wave functions calculated in a mean-field potential, with the correct quantum numbers and separation energy, and multiplied by a spectroscopic amplitude. The latter accounts for the fragmentation of single-particle strength due to beyond mean-field correlations. If one is not interested in the population of specific final states, but just in their sum, one may incorporate the  effect of this fragmentation by means  of a complex potential, whose imaginary part accounts for the spreading width of the single-particle levels into these more complicated configurations. This is the case of the dispersive optical potential, first introduced by  Mahaux and Sartor \cite{Mahaux86} and recently pursued by several groups (see Ref.~\cite{Dic17} for a recent review). The use of this dispersive potentials permits a natural extension of the inclusive breakup models to negative energies \cite{Uda89,Udagawa86}.   A recent work, using the IAV model
in prior form \cite{Potel:2015eqa}, has shown that this procedure leads to a
smooth transition between the positive and negative $E_x$ values and
hence between the breakup and transfer regions. However, the relation between the prior and form formulations for the case of transfer reactions with complex binding potentials has not been established to our knowledge. In particular, it remains to clarify the importance of the non-orthogonality term in this case. Indeed, for
real potentials, these results should lead to the well-known post-prior
equivalence used in transfer reactions, and the non-orthogonality term should not contribute in this case.

Guided by these considerations, in this paper, we address the post-prior equivalence for transfer
reactions of the form of $a+A\to b+ B^*$ in presence of complex $x+A$ potentials.
For that purpose, we revisit and generalize
the IAV model which allows us to describe the breakup ($E_x>0$) and transfer ($E_x<0$) regions in the same footing.
We will see that, in the extended version of the IAV model, the use of a complex  $U_{xA}$ potential
leads to different formulas for post and prior representations. As a practical application of the derived formulas, 
we present calculations for the $^{58}$Ni($d$,$pX$) reaction at $E=80$ MeV.

The paper is organized as follows. In Sec.~\ref{sec:theo} we summarize the main
formulas of the IAV model in post and prior forms, and outline its relation with the prior form UT model. We show how the IAV model can be naturally extended to final bound states. 
In Sec.~\ref{sec:calc}, the formalism is applied to the $^{58}$Ni($d$,$pX$) reaction. Finally, in Sec.~\ref{sec:conclu} we summarize the main results.

\section{Theoretical framework \label{sec:theo}}
In this section, we briefly summarize the IAV model, in its post and prior forms, and highlight its connection with the UT model. Further details can be found in Ref.~\cite{Ich86} as well as in our previous works  \cite{Jin15, Jin15b,Jin17}.
We write the process under study as
\begin{equation}
a(=b+x) + A \to b + B^* ,
\end{equation}
where the projectile $a$, composed of $b+x$, collides with the target
$A$, emitting the ejectile $b$ and leaving the residual system $B^*$ ($=x+A$) in any possible final state compatible with energy and momentum conservation. This includes $x+A$ states with  both positive and negative relative energies.

The IAV model, as well as the UT model, treats the particle $b$ as a spectator, meaning
that its interaction with the target nucleus is described with an optical
potential $U_{bA}$.

Using the post-form IAV model in DWBA, the inclusive breakup differential cross
section, as a function of the detected angle and energy of the fragment $b$, is given by

\begin{align}
\label{eq:dsdew}
\frac{d^2\sigma}{d\Omega_b E_b }\Big|^\mathrm{post} &=  \frac{2 \pi}{\hbar v_a} \rho(E_b) \sum_{c}
|\langle \chi^{(-)}_{b} \Psi^{c,(-)}_{xA} |V_\mathrm{post}| \chi_a^{(+)}
\phi_a  \phi^{0}_A \rangle |^2 \nonumber \\
& \times \delta(E-E_b-E^c) ,
\end{align}
where $v_a$ is the velocity of the incoming particle $a$,
$V_\mathrm{post} \equiv V_{bx} + U_{bA}-U_{bB}$ is the post-form transition operator,
$\rho_b(E_b)=k_b \mu_{b} /[(2\pi)^3\hbar^2]$ (with $\mu_b$ the reduced
mass of $b+B$ and $k_b$ their relative wave number), $|\phi_a\rangle$
and $|\phi^{0}_A\rangle$ are the projectile and target ground states,
$\chi_a^{(+)}$ and $\chi^{(-)}_{b}$  are distorted waves describing the
$a-A$ and $b-B$ relative motion by the optical potentials $U_{aA}$ and $U_{bB}$,
respectively, and $\Psi^{c,(-)}_{xA}$
is any possible state of the $x+A$ many body system, with $c=0$ denoting the $x$ and
$A$ ground states. Thus, the $c=0$ term in Eq.~(\ref{eq:dsdew}) corresponds to the processes in which the target remains in the ground state after the breakup, usually called {\it elastic breakup} (EBU), whereas the terms $c \neq 0$ correspond to the so-called non-elastic breakup 
(NEB) contributions.

The theory of IAV allows to perform the sum in a formal way, making use of the Feshbach projection formalism and the optical model reduction, and leading to  the following closed form for the NEB differential cross section:
\begin{equation}
\label{eq:iav}
\left . \frac{d^2\sigma}{dE_b d\Omega_b} \right |_\mathrm{IAV}^\mathrm{post} = -\frac{2}{\hbar v_{a}} \rho_b(E_b)
 \langle \psi_x^\mathrm{post} | W_x | \psi_x^\mathrm{post} \rangle   ,
\end{equation}
where $W_x$ is the imaginary part of the optical potential $U_x$,
which describes $x+A$ elastic scattering.  
$\psi_x^\mathrm{post}(\vecr_x)$ is a projected $x$-channel wave function which describes the $x-A$ relative motion for a given outgoing momentum $\vec{k}_b$ of the $b$ particle, and obtained after projection onto the $A$ ground state using the Feshbach formalism. It verifies the equation
\begin{equation}
\label{phix_post}
|\psi_x^\mathrm{post}\rangle  = G_x |\rho \rangle _\mathrm{post} ,
\end{equation}
where $G_x =1/(E_x - H_x+i\epsilon)$, with $H_x=T_x+U_x$, $E_x=E-E_b$ and 
$\langle \vec{r}_x |\rho_\mathrm{post}\rangle=(\chi_b^{(-)} \vec{r}_x| V_\mathrm{post}|\phi_a
\chi^{(+)}_{a}\rangle$.  
Udagawa and Tamura \cite{Uda81} proposed a very similar
formula for the same problem, but making use of the prior-form representation. The prior-form $x-$channel wave function reads
\begin{equation}
\label{phix_prior}
|\psi_x^\mathrm{prior}\rangle = G_x  |\rho \rangle _\mathrm{prior} ,
\end{equation}
where $\langle \vec{r}_x |\rho_\mathrm{prior}\rangle=(\chi_b^{(-)} \vec{r}_x| V_\mathrm{prior}|\phi_a
\chi^{(+)}_{a}\rangle$. Using this channel wave function, UT proposed the following prior-form inclusive breakup formula
\begin{align}
\label{eq:ut}
\left . \frac{d^2\sigma}{dE_b d\Omega_b} \right |_\mathrm{UT}= -\frac{2}{\hbar v_{a}} \rho_b(E_b)
 \langle \psi_x^\mathrm{prior} | W_x | \psi_x^\mathrm{prior} \rangle   ,
\end{align}which is analogous to the IAV formula, Eq.~(\ref{eq:iav}) .

Despite their formal analogy, the UT and IAV expressions give rise to different
predictions for NEB cross section. This discrepancy led to a long-standing dispute between these two groups. The problem has been also reexamined recently \cite{Pot15,Jin15}. These works have concluded that the UT formula is incomplete, and must be supplemented with additional terms, as we show below.  The comparison of the IAV and UT models with experimental data supports this interpretation \cite{Pot15,Jin15,Jin17,Moro2016,Carlson2016,Pot17}

In general, the NEB will contain also contributions coming from
the population of states below the breakup threshold of the $x+A$ system ($E_x <0$).
One would like to have a common framework to describe transfer to both  continuum
as well as bound states.  For that purpose, Udagawa and co-workers \cite{Uda89} proposed to extend the complex potential to negative energies.  Then, the bound states of the system are simulated by the eigenstates in this complex potential, whose imaginary part is associated with the spreading width of the single-particle
states. The latter  accounts for the fragmentation of these states into more complicated configurations
due to the residual interactions. The method has been recently reexamined by
Potel \textit{et al.} \cite{Potel:2015eqa}, who have provided an
efficient implementation of this idea. The key point is the realization that the Green function
$G_x (r_x , r_x')$ in Eqs.~(\ref{phix_post}) and (\ref{phix_prior}) can be
expressed for both $E_x>0$ and $E_x<0$ cases. Proceeding in this way, the application of
Eq.~(\ref{eq:iav}) to positive and negative energies is formally analogous.

The post-prior equivalence for transfer reactions with real binding potentials is well known in the
literature \cite{satchler83,Glendenning12}. However,  the post-prior equivalence
with complex binding potentials 
has never been investigated to our knowledge.

The relation between the post and prior formulae was first established  by IAV  \cite{Ich86}. 

To relate the post and prior inclusive breakup cross sections, we note that
\begin{align}
 V_\mathrm{post} & = V_\mathrm{prior} + (V_{bx}+T_{bx}+U_{aA}+T_{aA}) \nonumber \\
& - (U_{xA}+T_{xA}+U_{bB}+T_{bB}).
\end{align}
%

We consider a definite final state of the $x-A$ system, denoted $| \phi_n\rangle$ and evaluate the difference
\begin{widetext}
\begin{equation}
\label{eq:diffpsi}
\langle \phi_n |\psi_x^\mathrm{post}\rangle- \langle \phi_n |\psi_x^\mathrm{prior}\rangle=
\frac{ \langle  \phi_n \chi_b^{(-)}|(V_{bx}+T_{bx}+U_{aA}+T_{aA})-
(U_{xA}+T_{xA}+U_{bB}+T_{bB})  |\chi_a^{(+)}\phi_a \rangle }{E_x - H_x} .
\end{equation}
\end{widetext}

The first term in parenthesis can be replaced by $E$ when acting on  $|\chi_a^{(+)}\phi_a \rangle$. The second parenthesis, acting on  $\langle  \phi_n \chi_b^{(-)}|$ gives also the total energy $E$, provided $H_x$ is Hermitian (i.e. $U_x$ real). In that case, we have 
\begin{equation}
\langle \phi_n\chi_b^{(-)}|V_\mathrm{post}|\chi_a^{(+)}\phi_a\rangle = 
\langle \phi_n\chi_b^{(-)}|V_\mathrm{prior}|\chi_a^{(+)}\phi_a\rangle ,
\end{equation}
which corresponds to the well-known post-prior equivalence for transfer reactions. 

However, when $U_x$ is complex we can not perform the last step. Instead, we may rewrite  (\ref{eq:diffpsi}) as
\begin{align}
 \label{eq:psiNO}
 |\psi_x^\mathrm{post}\rangle-   |\psi_x^\mathrm{prior}\rangle&= \frac{(\chi_b^{(-)}|E_x-H_x |\chi_a^{(+)}\phi_a \rangle}{E_x-H_x} \nonumber \\
&=(\chi_b^{(-)}|\chi_a^{(+)}\phi_a \rangle ,
\end{align}
The function $|\psi_x^\mathrm{NO}\rangle=(\chi_b^{(-)}|\chi_a^{(+)}\phi_a \rangle$ is the so-called non-orthogonality (NO) overlap, also referred to in the literature as Hussein-McVoy term. Upon replacement of  this relation into Eq.~(\ref{eq:iav}) we finally get 
\begin{align}
\label{eq:3terms}
\frac{d^2\sigma}{d\Omega_b E_b }\Big|_\mathrm{IAV} =& \frac{d^2\sigma}{d\Omega_b E_b }\Big|_\mathrm{UT} +
\frac{d^2\sigma}{d\Omega_b E_b }\Big|_\mathrm{NO} + \frac{d^2\sigma}{d\Omega_b E_b }\Big|_\mathrm{IN}
\end{align}
where the first term is the UT prior-form formula of the NEB cross section, Eq.~(\ref{eq:ut}), 
\begin{align}
\left . \frac{d^2\sigma}{dE_b d\Omega_b} \right |_\mathrm{NO} = -\frac{2}{\hbar v_{a}} \rho_b(E_b)
 \langle \psi_x^\mathrm{NO} | W_x | \psi_x^\mathrm{NO} \rangle   ,
\end{align}
is the non-orthogonality term and 
\begin{align}
\left . \frac{d^2\sigma}{dE_b d\Omega_b} \right |_\mathrm{IN} = -\frac{4}{\hbar v_{a}} \rho_b(E_b)
 Re[\langle \psi_x^\mathrm{NO} | W_x | \psi_x^\mathrm{prior} \rangle ]  ,
\end{align}
is the interference term. 

Equation (\ref{eq:3terms}) shows that the IAV post-form formula and UT prior-form formula are not equivalent. The latter needs to be supplemented with two additional terms, which stem from the non-orthogonality of the initial and final states. Recent calculations comparing these expressions with experimental data support the interpretation of IAV  \cite{Jin15,Jin15b}.  

We also note that the relations (\ref{eq:psiNO}) and (\ref{eq:3terms}) hold for both $E_x<0$ and $E_x>0$ cases. That is, even in the transfer to bound states the prior form expression requires the inclusion of the non-orthogonality term. Only when the $x-A$ potential is real the contribution of these terms vanish in the DWBA expression.  It is one of the purposes of this work to assess the validity of (\ref{eq:3terms}) in actual calculations for complex $U_x$.

It is enlightening to consider the simple case in which $W_x$ is taken as a constant. In this case,  
if one inserts the Green's operator into Eq.~(\ref{eq:iav}), the double differential cross section for transferring particle $x$ to  bound states in post-form results
\begin{align}
\label{eq:trans_post}
\frac{\mathrm{d}^2\sigma}{\mathrm{d}E\mathrm{d}\Omega}\Big|^\mathrm{post}_\mathrm{IAV}&
=-\frac{2}{\hbar v_a} \rho_b(E_b)\langle  \rho |G_x^\dagger W_xG_x|  \rho   \rangle_\mathrm{post} 
\end{align}
Using the explicit form of $G_x$ and introducing the identity,
$\sum_n |\phi_n \rangle \langle \phi_n |$, we get 
\begin{equation}
\frac{\mathrm{d}^2\sigma}{\mathrm{d}E\mathrm{d}\Omega}\Big|_\mathrm{IAV}^\mathrm{post}
=\sum_n\mathcal{\omega}_n \frac{d\sigma_n}{d\Omega}  \Big|_\mathrm{post},
\end{equation}
where we have introduced the notation
\begin{equation}
\label{eq:wn}
\mathcal{\omega}_n=-\frac{1}{\pi}\frac{\Gamma_{n}}{(E_x-E_n)^2 +\Gamma_{n}^2},
\end{equation}
with $\Gamma_n=\langle \phi_n | W_x | \phi_n \rangle = W_x$
and 
\begin{align}
\label{eq:transfer}
\frac{d\sigma_n}{d\Omega} \Big|_\mathrm{post}& =\frac{2\pi}{\hbar v_a}\rho_b(E_b) \big|\langle \phi_n|\rho  \rangle_\mathrm{post}\big|^2 \nonumber \\
&= \frac{2\pi}{\hbar v_a}\rho_b(E_b)  |\langle  \phi_n \chi_b^{(-)}|V_\mathrm{post} |\chi_a^{(+)}\phi_a \rangle |^2 .
\end{align}
It should be noted that the single-differential cross section given 
by Eq.~(\ref{eq:transfer}) is nothing but the DWBA cross section of 
transfer cross sections. Equation (\ref{eq:wn}) shows that, for constant $W_x$, the energy distribution of the double differential cross section follows a Breit-Wigner (Lorentzian) shape for the bound  states $E_n<0$, and $\Gamma_n$ represents the spreading width of the single-particle levels generated by the $V_x$ potential. 

Notice that, in the limit case $W\to 0$, $\mathcal{\omega}_n \to \delta(E_x-E_n) $  and hence ${\mathrm{d}^2\sigma}/{\mathrm{d}E\mathrm{d}\Omega}|_\mathrm{post}$ is zero everywhere except at the pole energies  $E_x=E_n$.

\section{Calculations \label{sec:calc}}
We  consider the reaction $^{58}$Ni($d$,$pX$) at $E_{d}^\mathrm{lab}=80$~MeV. This reaction was also considered in our previous work \cite{Jin15}, where we compared with the inclusive breakup data from Ref.~\cite{Wu79} using  the original IAV model and considering only the $E_x>0$ region for the $n$-$^{58}$Ni residual system. Here, we extend these calculations to negative energies ($E_x<0$), studying the effect of the imaginary part of the $U_x$ potential, and comparing the post and prior results. 

In the present calculations, we consider for the $n-p$ interaction the simple Gaussian form of
Ref.~\cite{Auscdcc}. The deuteron and proton distorted waves are generated with
the same optical potentials used in Ref.~\cite{Jin15}. The neutron-$^{58}$Ni potential is extrapolated to negative energies by simply fixing its real and imaginary parts to their values at $E_n=1$~MeV for $E_n \leq 1$~MeV, that is, $U_x(E_x<1~{\rm MeV})=U_x(E_x=1~{\rm MeV})$.  
The bin procedure is
used to average the distorted wave $\chi_b$ over small momentum intervals to
evaluate the post-form formula. Although this is not required for the prior-form
formula, the same averaging procedure is adopted in that case for consistency with the post-form results. 

\begin{figure}[tb]
\begin{center}
 {\centering \resizebox*{0.9\columnwidth}{!}{\includegraphics{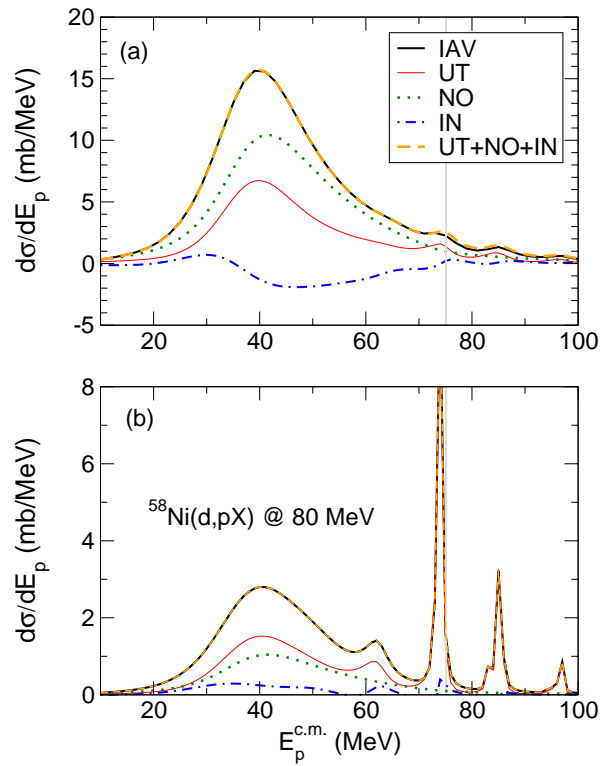}} \par}
\caption{\label{d58ni}(Color online)
(a) Angle-integrated proton energy spectra $^{58}$Ni($d$,$pX$) at $E_d=80$ MeV. The thick solid line is the post-form calculation (IAV model). The thin solid, dotted and dot-dashed lines are  the UT, NO, IN terms contributing to the prior-form cross section and  the dashed line is their sum. The vertical line indicates the threshold  ($E_n=0$) energy. (b) Same as panel (a), but with the imaginary part of the $n$-$^{58}$Ni reduced by a factor of 10. See text for details.}
\end{center}
\end{figure}

In Fig.~\ref{d58ni}(a) we present the post and prior calculations for the angle-integrated differential cross section of the outgoing protons as a function of their center-of-mass energy.  The black thick solid line is the post-form calculation obtained with the IAV post-form model and the red thin solid line is the UT prior-form calculation. It is seen that there is a significant difference between these two calculations,  for both $E_n<0$ and $E_n>0$ regions. When adding the NO (dotted line) and IN (dot-dashed) terms to the UT result, one obtains an excellent agreement with the IAV-post result both at positive and negative neutron energies. This demonstrates, for the first time to our knowledge, the post-prior equivalence of the transfer  cross section leading to bound states, in presence of complex binding potentials. This result has implications if, for example, a dispersive optical model potential is to be used to describe the bound states of the residual $B$ system \cite{Pot17}. 

The fact that the difference between the  IAV and UT results at negative neutron energies originates from the use of a  complex neutron potential  is illustrated in panel (b), where the imaginary part of this  potential is reduced by a factor of 10. As expected, for  $E_n>0$ this leads to  a reduction of the NEB cross section [c.f.~Eq.~(3)]. For $E_n<0$ the much weaker absorption leads to an almost perfect agreement between the IAV and UT formulas, which is the usual post-prior equivalence for transfer reactions. We note also that, for this weak-absorption case, the differential cross sections displays marked peaks at the position of the bound states and resonances of the neutron-$^{58}$Ni potential. In particular, a very narrow  $\ell=4$ resonance is found near the neutron threshold. Therefore, the role of the imaginary part is to increase NEB cross section, but also to smear the contribution of the bound states and resonances.  This is also apparent from Eq.~(\ref{eq:trans_post}) which, in the case of constant $W_x$, predicts a Lorentzian shape with a width given by $\Gamma=  W_x $. To conclude this section, we notice that, even in the limit of small $W_x$, the IAV and UT results differ for $E_n >0$. In this case, the addition of the NO and IN terms is essential to restore the post-prior equivalence, as shown in our previous work  \cite{Jin15b}.

\section{Summary and conclusions \label{sec:conclu}}
In summary, we have addressed the problem of the post-prior equivalence in the
calculation of inclusive transfer reactions of the form $A(a,b )B$, were $B$ is any bound state of the $x+A$ system. For that, we have considered  the post-form inclusive breakup model proposed by Ichimura, Austern and Vincent (IAV) \cite{Aus81,Ich86,Kas82}, conveniently extended to negative (bound states) of the $x+A$ system. We have also considered the prior-form model of Udagawa and Tamura (UT) \cite{Uda81}. 


We have shown that the  equivalence between the post-form (IAV) and prior-form (UT) expressions holds only for real $x-A$ potentials. For 
complex interaction, the non-orthogonality (NO) term is indispensable. Once this term is included, the post-prior equivalence is restored. 

To assess this equivalence at a numerical level, we  have performed calculations for the $^{58}$Ni($d$,$pX$) reaction at 80 MeV. We find that, when a complex potential is used for the $x-A$ system, the IAV and UT results significantly disagree, both for the unbound ($E_x>0$) and bound ($E_x<0$) regions. Inclusion of the NO term gives an excellent agreement between the post and prior cross sections.   We have also verified that, as the imaginary part of $U_x$ is reduced, the UT result approaches the IAV one, thus recovering the well-known post-prior equivalence of the DWBA formula.

We believe that the present results are relevant because they extend a fundamental property of the transition amplitude, namely, the post-prior equivalence, to the case of non-Hermitian binding potentials. In particular, the results will be useful in the context of the exclusive or inclusive transfer reactions with dispersive optical potentials, currently under development. 

\begin{acknowledgments}
We are grateful to Gregory Potel for his guidance in the extension of the IAV model to negative energies.
This work has been partially supported by the National Science Foundation
under contract. No. NSF-PHY-1520972 with Ohio University,
by the Spanish
 Ministerio de  Econom\'ia y Competitividad and FEDER funds under project
 FIS2014-53448-C2-1-P  and by the European Union's Horizon 2020 research and innovation program under grant agreement No.\ 654002.
\end{acknowledgments}

\bibliography{inclusive_prc.bib}
\end{document}